# TESTING COSMOGONIC MODELS WITH GRAVITATIONAL LENSING

Joachim Wambsganss*, Renyue Cen, Jeremiah P. Ostriker and Edwin L. Turner

Princeton University Observatory, Princeton, NJ 08544
* Max-Planck-Institut für Astrophysik, Karl-Schwarzschildstr.1, 85740 Garching, Germany

email:joachim@mpa-garching.mpg.de
(cen,jpo,elt)@astro.princeton.edu

Gravitational lensing provides a strict test of cosmogonic models. Detailed numerical propagation of light rays through a universe having a distribution of inhomogeneities derived from the standard CDM (cold dark matter) scenario with the aid of massive, fully nonlinear computer simulations is used to test the model. It predicts that we should have seen far more widely split quasar images than have been found. These and other inconsistencies rule out the COBE normalized CDM model with $\Omega = 1$ and $H_0 = 50$km/s/Mpc; but variants of this model might be constructed, which could pass the stringent tests provided by strong gravitational lensing.

Gravitational lensing *directly* measures fluctuations in the gravitational potential along lines of sight to distant objects. In contrast, the conventional tools for comparing theories with observations rely on either galaxy density or velocity information, both of which unavoidably suffer from the uncertainties with regard to density or velocity bias of galaxies over the underlying mass distribution, hampering our attempts to understand the more "fundamental" questions concerning the mass evolution and distribution. Thus, gravitational lensing provides a powerful independent test of cosmogonic models[1,2]. Each model for the development of cosmogonic structure (e.g. the HDM [hot dark matter] or CDM scenario) has at least one free parameter, the amplitude of the density (or potential) power spectrum. But now in the light of COBE observations[3], that parameter is fixed by the ($\pm 15\%$) determination on the $5^o - 10^o$ scale in the linear regime. With its amplitude fixed, a secure determination of the potential fluctuation on any scale provides a test; any single conflict between the theory and reality can falsify the former. The most leverage is obtained for tests made on scales as far as possible from the COBE measurements. The reason is that all models have an assumed power spectrum that passes through the COBE normalization point at the very large comoving scales ($\lambda \approx 1000$Mpc) fixed by that measurement. Since the slope of the power spectrum is a primary model dependent feature, the maximum variations amongst models occurs typically at the smallest scales. Thus one looks for tests at scales as small as possible, but they should not be so small as to be greatly influenced by the difficulty in modelling the physics of the gaseous, baryonic components ($\leq 10$kpc). Thus critical tests are best made on scales $0.01 \text{Mpc} < r < 1 \text{Mpc}$. The purpose of this paper is to use gravitational lensing from matter distributions on these scales to test the standard CDM scenario.

The model simulated here is the "standard" cold dark matter (CDM) scenario with $\Omega = 1$, $\lambda = 0$ and $H_0 = 100h = 50 \text{km/s/Mpc}$. Normalization, taken from the COBE first year results[3], corresponds to $\sigma_8 = 1.05^4$. In order to allow for the existence of very large-scale waves, we first ran an $L = 400h^{-1}\text{Mpc}$ size box with $500^3 = 10^{8.1}$ cells and $250^3 = 10^{7.2}$ particles. In addition, in order to have detailed small scale information we reran a total of 10 independent simulations with $L = 5h^{-1}\text{Mpc}$, having $500^3 = 10^{8.1}$ cells and $250^3 = 10^{7.2}$ particles. Knowing the distribution of overdensities on the $5h^{-1}\text{Mpc}$ scale from the large simulation, we can statistically convolve the small and large scale runs to produce simulated sheets or screens of matter spaced $5h^{-1}\text{Mpc}$ apart between the observer at $z = 0$ and a putative galaxy or quasar in the source plane at $z = z_s$. A large number of independent runs (ten runs were simulated) is required so that identical structures do not repeat along a line of sight. The details of the convolution method and tests of it using a high resolution P$^3$M simulation provided by Bertschinger and Gelb[5] will be presented in a subsequent paper. The method is statistically reliable for describing structures in the range 30kpc $< \Delta Lh < 1.2\text{Mpc}$, which corresponds roughly to splitting angles $5" < \theta < 200"$. On these scales we expect that dark matter dominates over baryons so that a dark matter only simulation is approximately valid.

A very preliminary attack on this problem has been presented in Paper I[2]. In that work no ray tracing was done. Rather we simply checked whether or not mass accumulations were greater than the critical level[6] at which multiple imaging will occur. In addition to the much better method used here, the convolution algorithm has been modified and improved over the one adopted in Paper I.

In our ray tracing routine we use the multiplane lens equations[7], and speed the calculation of deflection angles by use of the hierarchical tree code[8], with typically of order of 200-300 (grouped) screens used for each ray tracing. Amplitude on the image plane is simply given by differential area within a bundle of rays as compared to what it would have been had the propagation been through a universe with smoothly distributed matter. Figure (1) shows the amplitude distribution for sources of a redshift $z_s = 3$. When a given region on the source plane is reached by rays in separate disconnected regions of the image plane, the observer would see multiple images of the same object.

We have computed the distribution of magnifications for single and multiply-imaged point sources as a function of $z_s$, multiplicity of images, and distribution of angular splittings. In addition, for extended sources (mock galaxies), we have computed the expected shape distortions, frequency and properties of the giant arcs that would be seen, when the source are lensed by intervening clusters[9]. Figure 2 shows the probability of a splitting with separation of images greater than 5" and magnitude difference less than 1.5 mag as a function of source redshift. In fact "amplification bias"[6, 10, 11] will increase the probabilities over those shown

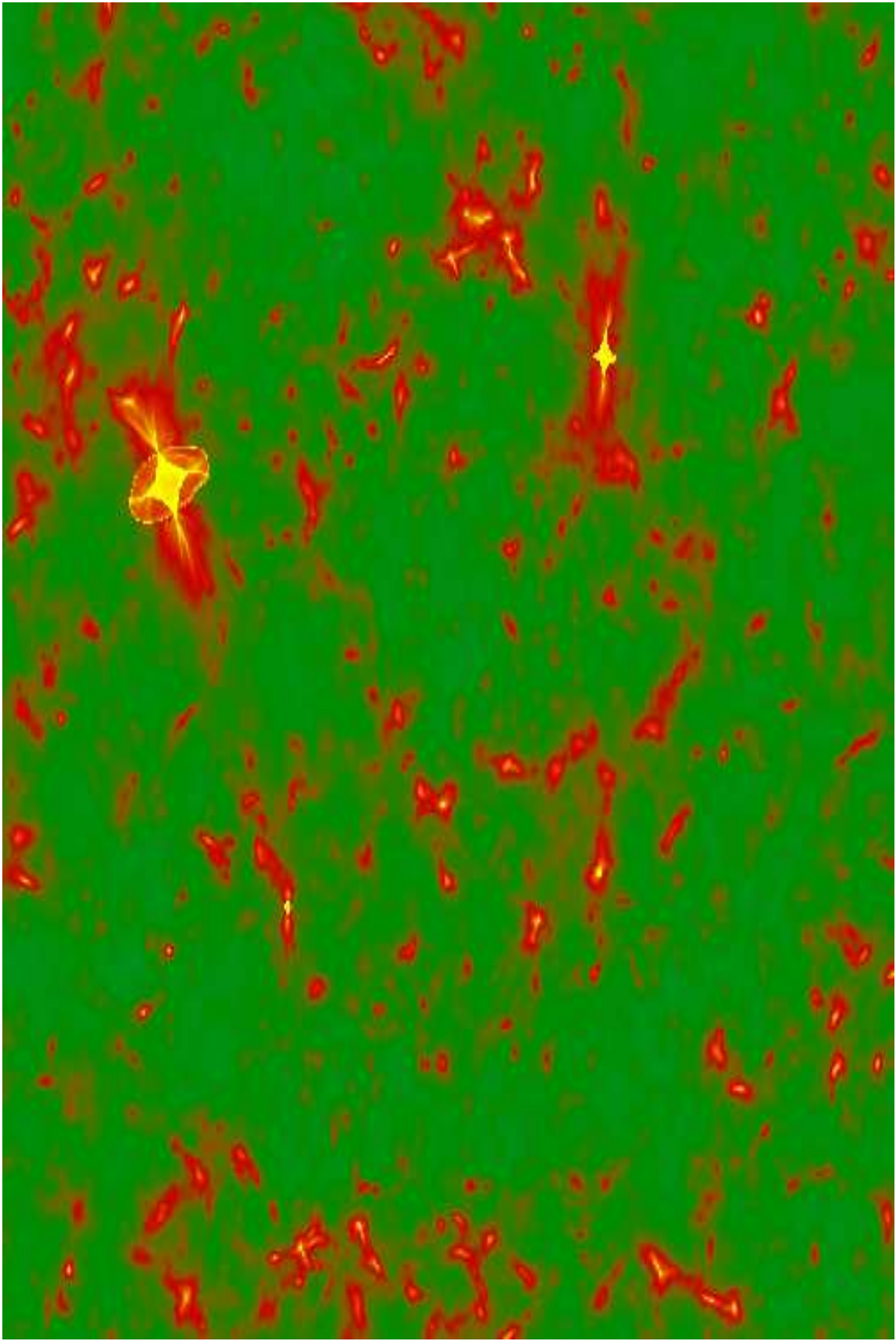

Figure 1. Example of the magnification due to the gravitational lens action of a CDM, $\Omega = 1$

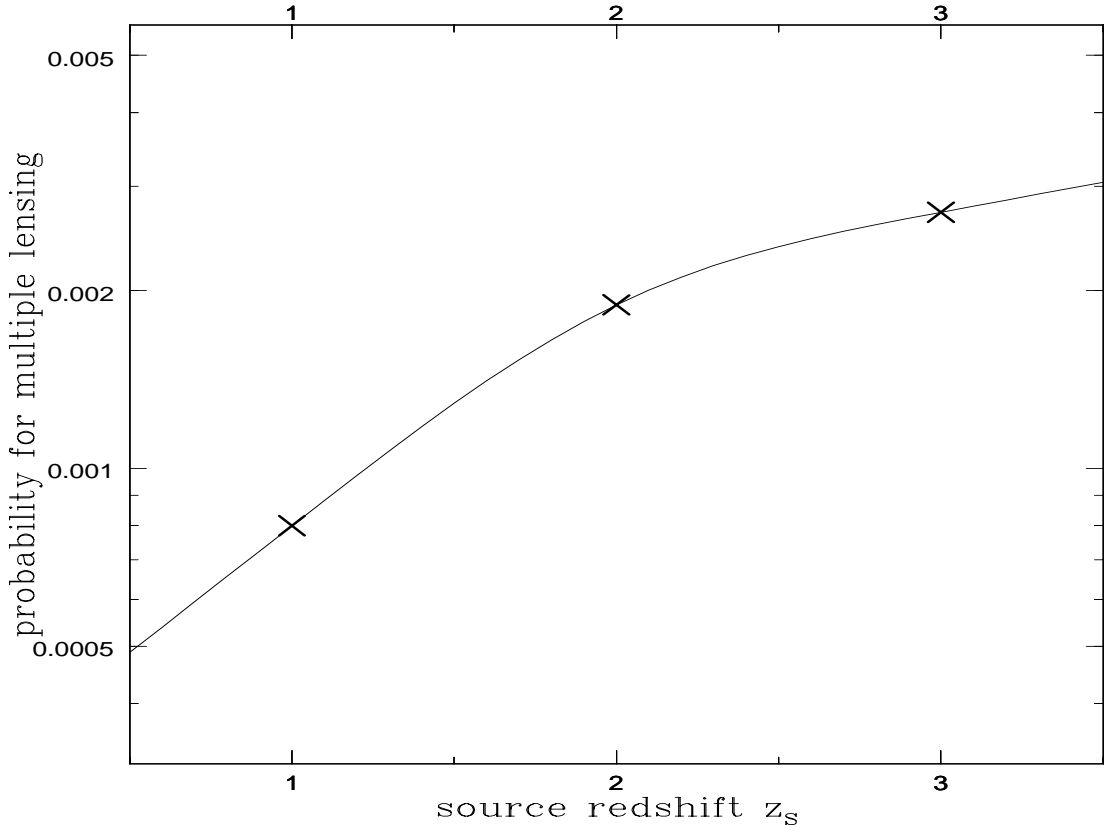

Figure 2

Probability of a splitting with separation of images greater than 5" and magnitude difference less than 1.5 as a function of source redshift.

in Figure 2 by a significant amount. Splittings larger than 5" should be common (when several thousand quasars have been examined), if this cosmogonic model were correct.

Probably the single most revealing statistic is the distribution of image separations expected for multiple sources as shown in Figure 3 for $z_s = 1$, 2 and 3. Notice that very large splittings should be the rule. Also revealing is the distribution of expected lens redshifts as shown in Figure 4. The lenses themselves should be close enough to be seen in almost all cases. On this issue the recent observation[12] of a lens candidate for the double quasar QSO2345+007 is extremely relevant. The separation of the two images is 7".06, the quasar redshift is $z_s = 2.15$ and the putative lens is at $z_l = 1.49$. We see from Figure 4 that, although 7 arcsecs separation can be produced in the CDM model, the probability that the lens is as far away as $z = 1.49$ is very small (2%) due to the relatively late formation of structure in this model. In open models structure formation occurs earlier.

It appears that all three of these results (shown in Figures 2,3,4) are seri-

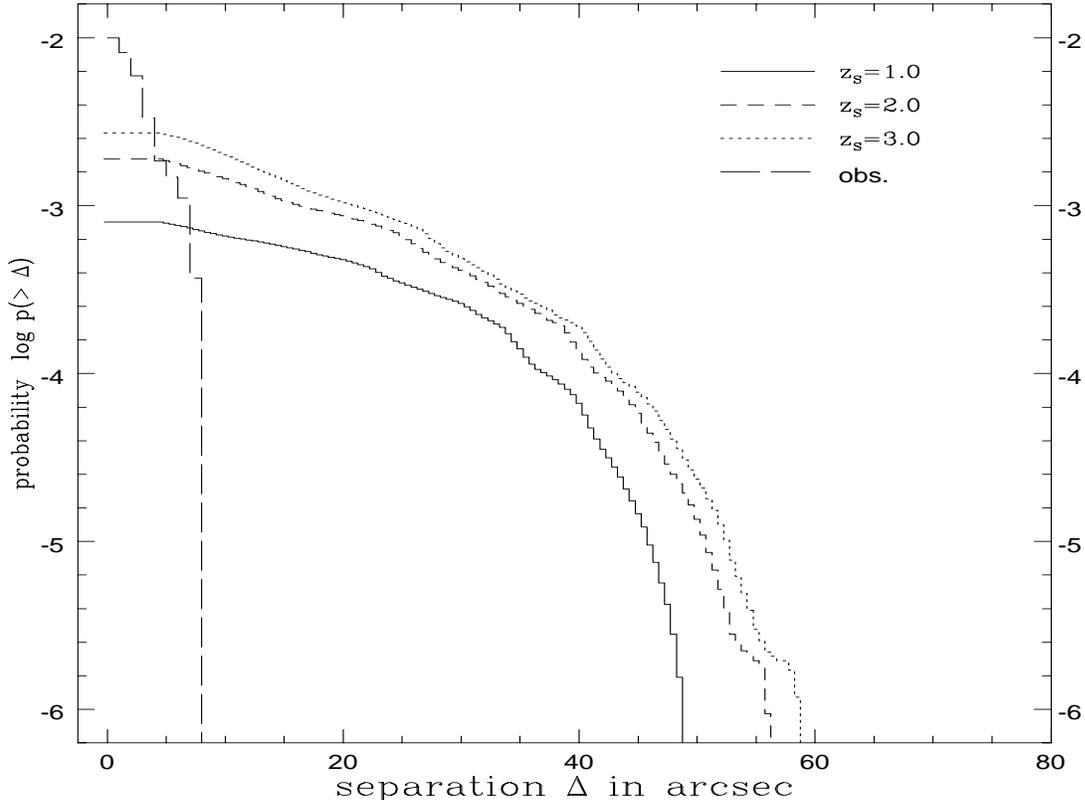

Figure 3

Multiple-lensing probability distribution as a function of image separations for sources at $z_s = 1, 2$ and $3$. Also shown as long dashed curve is the observed distribution[13].

ously in conflict with the existing observations. In particular, we find that the standard CDM model predicts that 0.0007 of all lines of sight to $z_s = 1$, 0.0014 of all lines of sight to $z_s = 2$ and 0.0020 of those to $z_s = 3$ will be multiply-imaged with angular splittings $\geq 10$" and amplification ratios of less than 1.5 magnitudes. Surveys[12-18] and occasional serendipitous discoveries have revealed 27 confirmed or possible multiply-imaged QSO's according to a recent compilation[19]. Detailed analysis[20,21] of the most statistically useful of these surveys[14,15,16] yields a lensing rate in the vicinity of a few tenth to one percent, consistent with the CDM predictions quoted above making allowance for plausible magnification biases[22]. However, as shown in Figure 3, all observed QSO lens systems have image splittings of less than 10", and the large majority, less than 5".

This sharply contradicts and thus falsifies the model. Since the large splitting, modest brightness ratio systems predicted by the model would be typically much easier to detect and recognize than those 27 which have actually been found, no escape by appeal to observational selection seems possible.

This failing of the model is not presented as an entirely new result, but

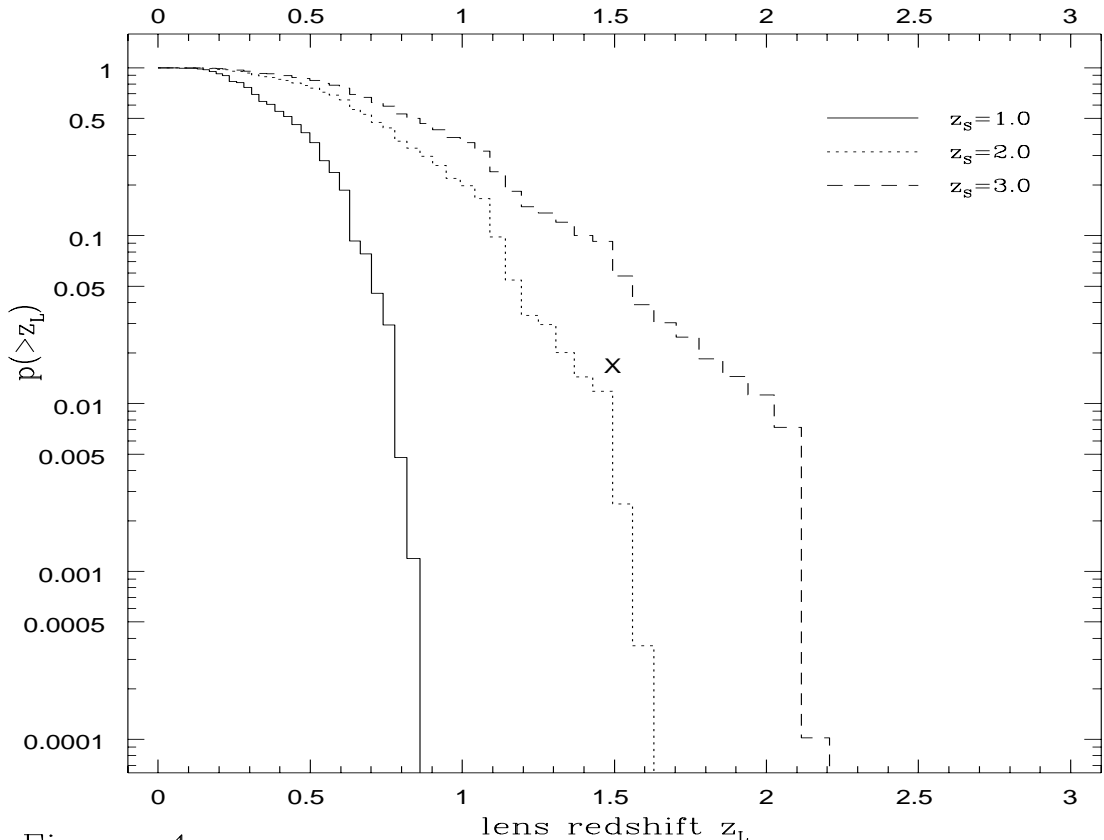

Figure 4

Integrated lensing probability distribution as a function of expected lens redshift. The X symbol indicates the recent observation[12] of of a lens candidate for the double quasar QSO 2345+007.

only as a new and more robust manifestation of a previously recognized problem, namely the excessively deep potential wells produced by the dark matter component in COBE normalized standard CDM[23, 24, 25]. These excessively deep potential wells lead both to excess galaxy pair-wise velocity dispersions and to the predicted excessive rate of large splitting lensing events. The virtue of the lensing test is that it is independent of other tests and is not subject to the same caveats concerning "bias" of galaxies with respect to dark matter.

Are there variant models that would not fail these tests? The simplest change that one can make in the scenario is to reduce the density of the clumped material in the universe, as it is the large number of large mass concentrations that produce the over-abundance of large splittings. Thus a lower $\Omega$ is clearly useful, but it would be premature to argue that the present results by themselves indicate $\Omega < 1$, as many other properties of the scenario ("temperature" of the dark matter, shape of the power spectrum etc) contribute to lensing properties. However the *directness* of gravitational lensing as a test for the growth of inhomogeneities, coupled with the rapidly increasing power of computers and numerical algorithms, makes one optimistic that calculations of the type reported on here should become a major tool for testing and discriminating among competing cosmological scenarios.

We thank J.R. Gott, C.S. Kochanek and P. Schneider for useful discussions, NCSA for our use of the Convex-3880 supercomputer, and NASA grants NAGW-2448 and NAGW-2173, NSF grants AST91-08103 and HPCC ASC-9318185 for financial support.

1. Narayan, R. & White, S.D.M. 1988, MNRAS, 231, 97p
2. Cen, R., Gott, J.R., Ostriker, J.P., & Turner, E.L. 1994, ApJ, 423, 1
3. Smoot, G.F., *et al.* 1992, ApJ(Letters), 396, L1
4. Efstathiou, G., Bond, J.R, & White, S.D.M 1992, MNRAS, 258, 1p
5. Bertschinger, E., & Gelb, J. 1994, ApJ, in press
6. Turner, E.L., Ostriker, J.P., & Gott, J.R. III 1984, ApJ, 284, 1 (TOG)
7. Schneider, P., Ehlers, J., & Falco, E.E., *Gravitational Lensing* (Springer Verlag, Berlin, 1992)
8. Wambsganss, J. Ph.D. Thesis, MPA-report 550 (1990)
9. Fort, B. & Mellier, Y. 1994, Astron. Astroph. Rev., 5, 239
10. Turner, E.L. 1980, ApJ(Letters), 243, L135
11. Narayan, R. & Wallington S. 1993, in: *Gravitational Lenses in the Universe*, Eds. J. Surdej et al., p.217 (Liège)
12. Fischer, P., Tyson, J.A., Bernstein, G.M., & Guhathakurta, P. 1994, ApJ(Letters), in press
13. Surdej, J. et al. 1993 AJ 105, 2064
14. Maoz, D. et al. 1993 ApJ 409, 28
15. Yee, H.K.C., Fillipenko, A.V., & Tang, D. 1993 AJ 105, 7
16. Crampton D., McClure, R.D., & Fletcher, J.M. 1992 ApJ 392, 23
17. Patnaik, A.R., Browne, I.W., Walsh, D., Chaffee, F.H., & Folts, C.B. 1992, MNRAS, 259, 1p
18. Hewitt, J.N. *et al.* 1989, Lec. Notes Phy., 330, 147
19. Surdej, J., & Soucail, G. 1993, in "Gravitational Lenses in the Universe", Proceedings of the 31st Liege International Astrophysical Colloquium, p205
20. Maoz, D., & Rix, H.-W. 1993, ApJ, 416, 425
21. Kochanek, C.S. 1993, ApJ, 417, 438; 419, 12
22. Fukugita, M., & Turner, E.L. 1991, MNRAS, 253, 99
23. Davis, M., Efstathiou, G., Frenk, C.S., & White, S.D.M. 1992, Nature, 356, 489
24. Bahcall, N.A., & Cen, R. 1992, ApJ(Letters), 398, L81
25. Ostriker, J.P. 1993, ARAA, 31, 689